\documentclass[letterpaper,11pt]{article}
\usepackage{graphicx}
\usepackage{comment}
\usepackage{url}

\begin{document}

\newcommand{\singlespace}{\renewcommand{\baselinestretch}{1} \normalsize}
\newcommand{\doublespace}{\renewcommand{\baselinestretch}{1.1} \normalsize}

\title{Accurate Estimation of Solvation Free Energy Using Polynomial Fitting
Techniques}

\author{Conrad Shyu\\
\url{conrads@uidaho.edu}\\
 \\
Department of Physics\\
University of Idaho\\
Moscow, ID 83844-0903\\
\and
F. Marty Ytreberg\\
\url{ytreberg@uidaho.edu}\\
 \\
Department of Physics\\
University of Idaho\\
Moscow, ID 83844-0903}

\maketitle \doublespace

%
\begin{abstract}
This report details an approach to improve the accuracy and precision of free
energy difference estimates using thermodynamic integration data (slope of the
free energy with respect to the switching variable $\lambda$) and its
application to calculating solvation free energy. The central idea is to
utilize polynomial fitting schemes to approximate the thermodynamic integration
data to improve the accuracy and precision of the free energy difference
estimates. In this report we introduce polynomial and spline interpolation
techniques. Two systems with analytically solvable relative free energies are
used to test the accuracy and precision of the interpolation approach (Shyu and
Ytreberg, \emph{J Comput Chem} 30: 2297--2304, 2009). We also use both
interpolation and extrapolation methods to determine a small molecule solvation
free energy. Our simulations show that, using such polynomial techniques and
non-equidistant $\lambda$ values, the solvation free energy can be estimated
with high accuracy without using soft-core scaling and separate simulations for
Lennard-Jones and partial charges. The results from our study suggest these
polynomial techniques, especially with use of non-equidistant $\lambda$ values,
improve the accuracy and precision for $\Delta F$ estimates without demanding
additional simulations. To allow researchers to immediately utilize these
methods, free software and documentation is provided via
\url{http://www.phys.uidaho.edu/ytreberg/software}.
\end{abstract}

%
\section{Introduction}
The Helmholtz or Gibbs free energy constitutes an important thermodynamic
quantity to understand how chemical species recognize each other, associate or
react. Examples of such problems include conformational equilibria and
molecular association, partitioning between immiscible liquids, receptor-drug
interaction, protein-protein and protein-DNA association, and protein stability
\cite{chipot07, reddy01}. Thermodynamic integration (TI) is a commonly used
approach for the calculation of free energy differences ($\Delta F$) between
two systems with potential energy functions $U_1$ and $U_0$, respectively
\cite{kirkwood35, mordasini00, shirts05, shirts03, ytreberg06}. The free energy
difference, $\Delta F = F_1 - F_0$, is equivalent to the reversible work to
switch from $U_0$ to $U_1$, and can be determined by estimating the integral
\begin{equation}
\Delta F = \int_{\lambda = 0}^{1} \left \langle \frac{\partial U _{\lambda}}
{\partial \lambda} \right \rangle _{\lambda} d \lambda.
\label{equ:integral}
\end{equation}
The notation $\left \langle \partial U _{\lambda} / {\partial \lambda} \right
\rangle _{\lambda}$ indicates an ensemble average at a particular value of
$\lambda$. The variable $\lambda$ permits continuously switching from one
energy function to another. Switching potential energies requires a
continuously variable energy function $U_{\lambda}$ such that $U_{\lambda = 0}
= U_0$ and $U_{\lambda = 1} = U_1$. In addition, the free energy function
$U_{\lambda}$ must be differentiable with respect to $\lambda$ for $0 \le
\lambda \le 1$ \cite{leech01}.

The relationship in Equation \ref{equ:integral} is exact. The numerical
estimation comes into play because the integral must be approximated by
performing simulations at various fixed, discrete, values of $\lambda$.
Typically, these discrete $\lambda$ values are used to convert the integral to
a sum (e.g., using quadrature). If the estimate of $\left \langle \partial U
_{\lambda} / {\partial \lambda} \right \rangle _{\lambda}$ is slow to converge,
then it is necessary to perform very long simulations in order reliably
estimate the average. In addition, $\left \langle \partial U _{\lambda} /
{\partial \lambda} \right \rangle _{\lambda}$ may heavily depend on $\lambda$
so that a large number of fixed $\lambda$ simulations is needed in order to
estimate the integral with sufficient accuracy.

Previously, Shyu and Ytreberg \cite{shyu09} reported the use of regression
techniques to calculate the free energy from TI data. Using analytical solvable
examples, the authors were able to achieve highly accurate estimates. We note
that regression does not produce a curve that goes through every data point
exactly. Instead the regression model only describes the tendency of the data
and does not represent the functional form.

The purpose of this current study is to introduce polynomial and spline
interpolation techniques to estimate $\Delta F$ and to use both interpolation
and regression methods to estimate a small molecule solvation free energy. This
differs from our previous study in that these polynomials interpolate the slope
of the free energy $dF / d \lambda = \left \langle \partial U _{\lambda} /
{\partial \lambda} \right \rangle _{\lambda}$ as a function of $\lambda$. The
derived curve is forced to pass through each data point. Regression, by
contrast, constructs the curve that minimizes the errors between the data and
extrapolated points. The key motivation for this study is that, even if the
averages $\left \langle \partial U _{\lambda} / {\partial \lambda} \right
\rangle _{\lambda}$ are fully converged (i.e., infinitely long sampling), there
will be error in the $\Delta F$ estimates due to the fact that one must
estimate the integral. Here we present methods which reduce this error and
construct the polynomial that represents the functional form of TI data. We
also examine the use of both equidistant and non-equidistant $\lambda$ values.
Two test systems previously used by Shyu and Ytreberg \cite{shyu09} with
analytical $\Delta F$ values were utilized to quantify the accuracy and
precision of the interpolation techniques. We also used both regression and
interpolation techniques to estimate a small molecule solvation free energy.
The results from our simulations suggest that polynomial fitting, especially
with use of non-equidistant $\lambda$ values, improves the accuracy and
precision for $\Delta F$ estimates without demanding additional simulations. In
addition, polynomial fitting techniques permit accurate estimates of free
energy difference from TI without the need for soft-core scaling and separate
simulations for partial charges and Lennard-Jones potentials.

%
\section{Theory}
The primary focus of this study is to present a mathematical framework for the
analysis of fixed TI simulation data using polynomial and spline interpolation
techniques. The objective is then to construct a polynomial that best fits the
TI data. The functional form of the simulation data is represented by a series
of data points $\left \{ \lambda, dF / d \lambda \right \}$, thus a polynomial
is constructed through these data. Interpolation refers to the problem of
determining a function that exactly represents given data points.
Mathematically the polynomial obtained from interpolation is considered to be
the one that best represents all data points within the given interval $\left (
\lambda = [ 0, 1 ] \right )$ \cite{faires93, henry81}.

Polynomial interpolation and regression techniques generally give a similar
result, especially when high degrees of polynomials are employed to fit the TI
data. However, the curve obtained from polynomial interpolation passes though
each data point whereas the curved generated via regression does not. If TI
simulations are fully converged (i.e., infinite sampling), then polynomial
interpolation should give the most accurate and precise estimate. Regression,
on the other hand, utilizes the least squares method that minimizes the sum of
the square of the difference between the actual value and the approximated one.
Regression does have an advantage over interpolation in that the approximation
is independent of the number of data points. However, we note that regression
may not give the best estimates if the TI data include several inflection
points due to the fact that regression does not attempt to construct a curve
that passes through every data point.

%
\subsection{Mathematical Notation}
To simplify the mathematical expressions and to be consistent with the most
commonly used notation in polynomial and spline interpolation papers, we denote
the switching variable $\lambda$ by $x$, and, $dF / d \lambda$ by $y$. We also
denote the actual continuous function $dF / d \lambda$ by $f \left( x \right)$.
The functional form of the simulation data is thus represented by a series of
data points $\left \{ \lambda, dF / d \lambda \right \} = \left \{ x, y \right
\}$, and the approximating function $p ( x )$ is constructed through these data
points. The following sections briefly introduce mathematical definitions and
properties of both Lagrange and Newton polynomials, and the cubic spline
function.

%
\subsection{Lagrange and Newton Polynomial Interpolation}
The classical Weierstrass theorem establishes the mathematical foundation for
polynomial approximation \cite{bishop61}. The theorem asserts that there exists
a polynomial $p \left( x \right)$ for approximating the continuous function $f
\left( x \right)$ defined within a closed interval $\left[ a, b \right] =
\left[ 0, 1 \right]$, and the polynomial approximation $p \left( x \right)$ can
get arbitrarily close to the function $f \left( x \right)$ as the degree of
polynomial is increased \cite{jeffreys88}. The most straightforward method to
obtain the polynomial $p \left( x \right)$ of degree $n$ is to calculate the
values of $f \left( x \right)$ for $n + 1$ distinct fitting data points within
the interval of $a \le x \le b$, and satisfy the simultaneous equation $y_i = f
\left( x_i \right) = p \left( x_i \right)$, for $i = 0, 1, 2, \ldots, n$.
Lagrange and Newton polynomials are the most commonly used methods for
interpolation \cite{jeffreys88, werner84}.

The coefficients of the above polynomial $p \left( x \right)$ then form the
basis of the Lagrange interpolation formula. The Weierstrass theorem contends
that there is always a unique polynomial $p \left( x \right)$ of degree $n$
that satisfies these conditions. The Lagrange interpolation polynomial $L
\left( x \right)$ is a linear combination of Lagrange basis polynomials $l_i
\left( x \right)$ such that
\begin{equation}
L \left( x \right) = \sum_{i = 0}^n y_i l_i \left( x \right) = \sum_{i = 0}^n
y_i \left[ \prod \limits_{i \ne j, j = 0}^n \frac{x - x_j}{x_i - x_j} \right].
\end{equation}
It is worth noting that the degree of polynomial $n$ cannot be chosen
arbitrarily but is determined by the number of data points $n + 1$. This is
important since the degree of interpolating polynomials will thus be determined
by the number of $\lambda$ values used to estimate $\Delta F$.

Mathematically, the Lagrange form of the interpolation polynomial is generally
preferred in proof and theoretical arguments because the derivatives of the
polynomial are continuous, and maxima or minima always exist \cite{faires93,
henry81}. The construction of Lagrange polynomials, however, is computationally
demanding because all Lagrange basis polynomials have to be reevaluated each
time $y_i$ is updated. By contrast, Newton interpolation utilizes the divided
difference (more on this below) which is generally more computationally
efficient \cite{lee89}.

A more practical form of interpolation polynomial for computational purposes is
given by Newton polynomials that utilize the divided difference. The divided
difference is defined as the ratio of the difference in the function values,
$y_i$ or $f \left( x_i \right)$, at two points divided by the difference in the
values of the corresponding independent variable, $x_i$. Divided differences do
not require the recalculations of coefficients if new data points are included.
As a consequence, it is generally more computationally efficient to use the
divided differences method for interpolation \cite{werner84}. Similar to the
Lagrange polynomial, the Newton polynomial is a linear combination of basis
polynomials
\begin{equation}
N \left( x \right) = \sum \limits_{i = 0}^n a_i \left[ \prod \limits_{j = 0}^{i
- 1} {\left( x - x_j \right)} \right],
\label{equ:divide} \end{equation}
where the coefficients $a_i = g \left[ {x_0, \ldots, x_i} \right]$ is the
notation for divided differences. The first divided difference, for example,
between data points $x_0$ and $x_1$ is given by
\begin{equation}
a_1 = g \left[ x_0, x_1 \right] = \frac{f(x_1) - f(x_0)}{x_1 - x_0} = \frac{y_1
- y_0}{x_1 - x_0}.
\end{equation}
The second divided difference between data points $x_0$, $x_1$, and $x_2$ is
given by
\begin{equation}
a_2 = g \left[ x_0, x_1, x_2 \right] = \frac{g [x_1, x_2] - g [x_0, x_1]}{x_2 -
x_0}.
\end{equation}
Accordingly the $n$-th divided difference between data points $x_0, x_1,
\ldots, x_n$ is given by
\begin{equation}
a_n = g \left[ x_0, x_1, \ldots, x_n \right] = \frac{g [x_1, x_2, \ldots, x_n]
- g [x_0, x_1, \ldots, x_{n - 1}]}{x_n - x_0}.
\label{equ:differ} \end{equation}
The Newton interpolation polynomial is then given by
\begin{eqnarray}
N \left( x \right) &=& f(x_0) + g[x_0, x_1](x - x_0) +
g[x_0, x_1, x_2](x - x_0)(x - x_1) + \nonumber \\
& & \ldots + g[x_0, x_1, \ldots, x_i](x - x_0)(x - x_1) \ldots (x - x_{i - 1}).
\end{eqnarray}
The calculations of the divided differences form a successive, recurrent
relationship between the previous two coefficients. Computationally, the
divided differences can be written in the form of a table that significantly
simplifies the algorithmic implementation \cite{werner84}.

%
\subsection{Spline Function Interpolation}
Studies have suggested that the spline function is typically preferred over
polynomials because the interpolation error can be reduced by using low degrees
of polynomials \cite{ahlberg67, greville69, schultz70}. Spline interpolation
also avoids the Runge's phenomenon which occurs when using high degree
polynomials \cite{epperson87}. Runge's phenomenon is when approximation errors
increase dramatically as the degree of interpolation polynomials increases.

Spline functions are piecewise polynomials of the same degree \cite{greville69,
deboor78}. Unlike the Lagrange and Newton interpolation polynomials, the degree
of the piecewise polynomials does not depend on the number of data points. The
piecewise polynomials join together at the data points $\left \{ \lambda, dF /
d \lambda \right \}$ called knots and must be continuous \cite{faires93}. The
main advantages of spline interpolation are stability and simplicity
\cite{wold74}. Moreover, the use of polynomials of lower degree offers the
possibility to avoid the oscillatory behavior that commonly arises from fitting
a single polynomial exactly to a large number of data points \cite{faires93}.

We utilize the cubic spline which is the most commonly used type. A cubic
spline function is a set of third degree piecewise polynomials that provide a
smooth curve passing through all data points. Because the segments join with
matching derivatives up to order two, the curvature of the polynomials changes
smoothly along the knots. The fundamental idea behind cubic spline
interpolation is based on the engineer's tool used to draw smooth curves
through a number of points. This spline consists of weights attached to a flat
surface at the points to be connected. A flexible strip is then bent across
each of these weights, resulting in a smooth curve. The mathematical spline is
similar in principle. The points, in this case, are TI data. The weights are
the coefficients of the cubic polynomials used to interpolate the data
\cite{schultz73}.

Mathematically, the cubic spline function $S_i \left( x \right)$ is defined as
\begin{eqnarray}
S_i \left( x \right) &=&
\frac{z_{i + 1}}{6h_i} \left( x - x_i \right) ^3 +
\frac{z_i}{6h_i} \left( x_{i + 1} - x \right) ^3 + \nonumber \\
& & \left( \frac{y_{i + 1}}{h_i} - \frac{h_i}{6} z_{i + 1} \right)
\left( x - x_i \right) + \left( \frac{y_i}{h_i} - \frac{h_i}{6} z_i \right)
\left( x_{i + 1} -x \right),
\label{equ:factor} \end{eqnarray}
where $h_i = x_{i + 1} - x_i$ is the size of the $i$-th interval $\left[ x_i,
x_{i+1} \right]$. The spline function $S_i \left( x \right)$ provides a cubic
polynomial on the interval $x \in \left[ x_i, x_{i + 1} \right]$. The values of
$z_i$ are the second derivatives at the the endpoints such that $z_i = S''_i
\left( x_i \right)$. To compute the values of $z_i$, it is necessary to solve
the recurrent equation
\begin{equation}
h_{i - 1} z_{i - 1} + 2 \left( h_{i - 1} + h_i \right) z_i + h_i z_{i + 1} =
6 \left( \frac{y_{i + 1} - y_i}{h_i} - \frac{y_i - y_{i + 1}}{h_{i - 1}}
\right),
\end{equation}
for $i = 0$ to $n - 1$.

In the current study, natural spline boundary conditions were implemented to
construct the interpolation polynomials. There are two boundary condition types
which could be utilized. The natural cubic spline imposes the boundary
conditions $z_0 = 0$ and $z_n = 0$, and the clamped cubic spline requires $z_0$
and $z_n$ for a given function \cite{ahlberg67, greville69}. When the natural
boundary conditions are used, the spline assumes the shape that a long flexible
curve would take if it is forced to go through all the data points. A natural
spline permits the slope at the ends be free to equilibrate to the position
that minimizes the oscillatory behavior of the curve. A clamped cubic spline,
on the other hand, further requires a piecewise cubic function which passes
through the given set of knots with a given function or value. The derivatives
and second derivatives of adjacent cubic polynomials must also agree at the
interior abscissae. It is worth noting that spline interpolation polynomials
are piecewise and do not always approximate the $dF / d \lambda$ completely.
This is particularly evident when the choice of $\lambda$ neglects the
inflection points along the curve.

%
\subsection{Algebraic Formulation}
To minimize the numerical errors in the $\Delta F$ estimation process,
interpolating polynomials were first transformed into their analytical forms,
and then all integrals were evaluated analytically. The derivation of the
analytical integration form for the interpolating polynomials involves the
expansion of every coefficient in the polynomial equation. The rearrangement of
coefficients permits analytical integration of the interpolating polynomials.
The analytical integral for the Lagrange polynomial is expressed as
\begin{equation}
\int L \left( x \right) dx = \sum_{i = 0}^n y_i \int l_i \left( x \right) dx =
\sum_{i = 0}^n y_i \int \prod_{i \ne j, j = 0}^n \frac{x - x_j}{x_i - x_j} dx.
\end{equation}
Similarly, the analytical integral for the Newton polynomial is expressed as
\begin{equation}
\int N \left( x \right) dx = \sum_{i = 0}^n a_i \int \prod_{j = 0}^{i - 1}
\left( x - x_j \right) dx,
\end{equation}
where the coefficient $a_i$ is the notation for divided differences; see
Equation \ref{equ:divide}.

For spline functions, analytical integrations must be performed at each
subinterval. Similar to that of Lagrange and Newton polynomial, the integration
form requires the expansion and rearrangement of all coefficients. The
analytical integral form of the cubic spline is expressed as
\begin{eqnarray}
\int S_i \left( x \right) dx
&=& x^3 \left[ \frac{1}{6h_i} \left( z_{i+1} - z_i \right) \right] +
x^2 \left[ \frac{1}{2h_i} \left( z_i x_{i+1} - z_{i+1} x_i \right) \right] +
\nonumber \\
& & x \left[ \frac{1}{2h_i} \left( z_{i+1} x_i^2 - z_i x_{i+1}^2 \right) +
\frac{1}{h_i} \left( y_{i+1} - y_i \right) -
\frac{h_i}{6} \left( z_{i+1} - z_i \right) \right] + \\
& & \frac{1}{6h_i} \left( z_i x_{i+1}^3 - z_{i+1} x_i^3 \right) +
\frac{1}{h_i} \left( y_i x_{i+1} - y_{i+1} x_i \right) +
\frac{h_i}{6} \left( z_{i+1} x_i - z_i x_{i+1} \right). \nonumber
\end{eqnarray}

Similar to that of Lagrange and Newton interpolation polynomials, each $x_i$ is
replaced by $\lambda_i$ and $y_i$ by $\left \langle \partial U _{\lambda} /
\partial \lambda \right \rangle _{\lambda}$. Specifically, for Lagrange
interpolation
\begin{equation}
\Delta F = \int \limits_0^1 L \left( \lambda \right) d \lambda = \sum_{i = 0}^n
\left \langle \frac{\partial U _{\lambda}}{\partial \lambda} \right \rangle
_{\lambda _i} \int \limits_0^1 \prod_{i \ne j, j = 0}^n \frac{\lambda - \lambda
_j}{\lambda_i - \lambda _j} d \lambda \ \ \ \textrm{(Lagrange)}
\label{equ:lagrange}
\end{equation}
and for Newton interpolation
\begin{equation}
\Delta F = \int \limits_0^1 N \left( \lambda \right) d \lambda = \sum_{i = 0}^n
a_i \int \limits_0^1 \prod_{j=0}^{i-1} \left( \lambda - \lambda_j \right) d
\lambda \ \ \ \textrm{(Newton)}
\label{equ:newton}
\end{equation}
with the divided difference $a_i = g \left[ \lambda _0, \lambda _1, \ldots,
\lambda _i \right]$ given by Equation \ref{equ:differ}. To calculate $\Delta F$
using cubic spline, integrals are evaluated at each subinterval $\left [
\lambda _{i - 1}, \lambda _i \right ]$, and the total $\Delta F$ is the sum of
all integrals
\begin{equation}
\Delta F = \sum_{i = 1}^{n} \int_{\lambda _{i - 1}}^{\lambda _i} S_i \left(
\lambda \right) d \lambda \ \ \ \textrm{(Cubic Spline)}
\label{equ:spline}
\end{equation}
where $S_i$ are given by Equation \ref{equ:factor}.

%
\section{Computational Details}
Two types of free energy calculations were performed in order to test our
interpolation approaches. The first type utilized two test systems with
analytical $\Delta F$ solutions \cite{shyu09}. Simulations were performed using
the same protocols as reported by Shyu and Ytreberg \cite{shyu09}. These test
systems provide an unbiased measure of the accuracy and precision of the
interpolation techniques for estimating $\Delta F$.

The second type of simulation was to compute the solvation free energy for a
4-hydroxy-2-butanone (BUQ) molecule \cite{burkhard00}. We used both regression
\cite{shyu09} and interpolation techniques and compared them to use of
quadrature. Details of the solvation free energy calculation are described
below.

%
\subsection{Test Systems}
Procedures for the simulations of the two test systems were previously
described by Shyu and Ytreberg \cite{shyu09}. Briefly, these systems involve
two potential functions $U_0 \left( \xi \right) = \frac{1}{2} \xi ^2$ and $U_1
\left( \xi \right) = 2 \left( \xi - 5 \right) ^2$ for system one, and $U_0
\left( \xi \right) = \frac{5}{2} \xi ^2$ and $U_1 \left( \xi \right) =
\frac{1}{2} \left( \xi - 5 \right) ^2$ for system two with the analytical free
energy $\Delta F = - \frac{1}{2} \ln \frac{1}{4}$ and $\Delta F = - \frac{1}{2}
\ln 5$ respectively \cite{shyu09}. An equal amount of simulation time
(1,000,000 Monte Carlo steps) was spent for each of six ($\lambda = 0.0, 0.2,
0.4, 0.6, 0.8$, and $1.0$) and eleven ($\lambda = 0.0, 0.1, 0.2, \ldots, 0.9$,
and $1.0$) equidistant $\lambda$ values. Identical equilibrium simulations were
performed on the corresponding six non-equidistant $\lambda$ values ($\lambda$
= 0.0, 0.0955, 0.3455, 0.6545, 0.9045, and 1.0) and eleven ($\lambda$ = 0.0,
0.0245, 0.0955, 0.2061, 0.3455, 0.5, 0.6545, 0.7939, 0.9045, 0.9755, and 1.0).
Averages of the slope $dF / d \lambda = \left \langle
\partial U _{\lambda} /
\partial \lambda \right \rangle _{\lambda}$ were collected for each value of
$\lambda$. For each $\lambda$ value, simulations were given 1,000 steps to
equilibrate initially, then were allowed to proceed for 1,000,000 Monte Carlo
steps. Each trial started with an arbitrarily chosen initial position for the
particle. Simulations were performed sequentially starting at $\lambda = 0$.
Trial moves for Metropolis Monte Carlo \cite{metropolis53} were generated by
adding a uniform random deviate between -0.5 and 0.5 to the current position
resulting in a 40 to 50\% acceptance ratio. A total of 1,000 independent trials
were generated from each case. We recorded the biases ($\Delta F
_{\textrm{exact}} - \Delta F _{\textrm{estimate}}$) for each of 1,000
independent runs and the corresponding averages and standard deviations were
calculated in order to provide a measure of the accuracy and precision of the
$\Delta F$ estimates.

%
\subsection{BUQ Solvation}
We applied our interpolation technique to compute the solvation free energy for
a BUQ molecule \cite{burkhard00}. We first obtained a reference estimate for
the solvation free energy $\Delta F _{\textrm{ref}}$. To obtain $\Delta F
_{\textrm{ref}}$ with high accuracy, we performed separate simulations for
changes in the Lennard-Jones parameters and the partial charges
\cite{steinbrecher07}. Studies have shown that use of separate simulations
provides a higher accuracy estimate and avoids possible endpoint singularity
problems \cite{shirts03, steinbrecher07, zacharias94}. Thus, the first stage to
obtaining $\Delta F _{\textrm{ref}}$ was to compute the free energy associated
with the Lennard-Jones potential. This was accomplished by setting all the
partial charges of the solute to zero. We then ``grew'' the BUQ molecule in the
solvent using soft-core scaling to ensure a smooth potential energy curve, even
for interpenetrating molecules \cite{zacharias94, beutler94}. Once the neutral
atoms are fully grown in the solvent, we performed the second stage to compute
the free energy associated with the partial charges. This was accomplished by
increasing the partial charges from zero to their final values given by the
forcefield. For both stages, we employed trapezoidal quadrature to integrate
the free energy slope and obtain the free energy for each of the two stages. We
computed $\Delta F _{\textrm{ref}}$ using this two-stage approach for both
inserting and deleting BUQ.

The initial coordinates for the BUQ were extracted from the FKBP-ligand
complexes (1D7J) \cite{burkhard00} which were retrieved from the Protein Data
Bank. The topologies for BUQ were then generated by the PRODRG server
\cite{schuttelkopf04}. The simulations were performed using the GROMACS package
3.3.3 \cite{lindahl01} at constant temperature in a periodic cubic box with an
edge length of approximately 2.4 nm and the default GROMOS-96 43A1 forcefield.
This corresponded to approximately 500 SPC \cite{robinson96} water molecules.
The neighbor list for nonbounded interactions was updated every 10 steps. The
bond distances and bond angles of water were constrained using the LINCS
algorithm \cite{hess97}. To obtain a isothermal-isobaric ensemble at 300 K, a
leap-frog stochastic dynamics \cite{gunsteren88} was used to integrate the
equations of motion with a 2.0 fs timestep. The temperature was maintained
using Langevin dynamics with a friction coefficient of 1.0 amu/ps. The pressure
was maintained at 1.0 atom using the Parrinello-Rahman algorithm
\cite{parrinello80}. The coupling time was set to 1.0 ps, and the isothermal
compressibility was set to $4.5 \times 10^{-5}$ bar$^{-1}$. Particle-mesh
summation Ewald algorithm \cite{darden93, essmann95} was used for
electrostatics with a real-space cutoff of 1.0 nm and a Fourier spacing of 0.1
nm. Van der Waals interactions used a cutoff with a smoothing function such
that the interactions smoothly decayed to zero between 0.75 nm and 0.9 nm.
Dispersion corrections for the energy and pressure were also utilized
\cite{allen89}.

The BUQ system was first minimized using steepest descent for 500 steps. 100 ps
of constant volume simulation were performed for system equilibration, followed
by 100 ps of constant pressure simulation. Then 10 ns constant pressure
simulations were preformed with eleven equidistant $\lambda$ values ($\lambda =
0.0, 0.1, 0.2, \ldots, 0.9$, and $1.0$). Simulations were performed
sequentially starting at $\lambda = 0$ and averages of the slope $dF / d
\lambda = \left \langle \partial U _{\lambda} / \partial \lambda \right \rangle
_{\lambda}$ were collected for each value of $\lambda$. For the first stage, we
computed the free energy for the Lennard-Jones interactions using a soft-core
scaling parameter of $\alpha = 0.5$. For the second stage, we computed the free
energy for the partial charge interactions using $\alpha = 0.0$
\cite{steinbrecher07}. Four additional $\lambda$ values ($\lambda = 0.65, 0.75,
0.85, 0.95$ for deleting BUQ and $\lambda = 0.05, 0.15, 0.25, 0.35$ for
inserting) were included to the simulations for Lennard-Jones potentials to
further refine the $\Delta F$ soft-core estimate where the free energy slopes
undergo the most change. The total free energy is the sume of the free energies
from simulations for the Lennard-Jones and partial charges. We determined that
$\Delta F ^{\textrm{del}} _{\textrm{ref}} = 31.6$ kJ/mol for deleting BUQ and
$\Delta F ^{\textrm{ins}} _{\textrm{ref}} = -31.5$ kJ/mol for inserting.

%
\section{Results and Discussion}
We now summarize the results from our polynomial fitting methods applied to two
test systems, and to BUQ solvation. The analytical test systems provide an
objective assessment of the accuracy and precision of the $\Delta F$ estimates.
The BUQ solvation provides a demonstration of our approach on a common chemical
problem. For the results below, interpolation polynomials for each set of TI
data were first transformed into the analytical form (Equations
\ref{equ:lagrange}, \ref{equ:newton}, and \ref{equ:spline}) and integration was
then performed analytically. Results presented below show that use of
interpolation with non-equidistant $\lambda$ values results in accurate and
precise $\Delta F$ estimates.

%
\subsection{Test Systems}
Table \ref{tab:analytical} summarizes the averages and standard deviations of
biases for system one and two from 1,000 independent trials using six and
eleven equidistant and non-equidistant $\lambda$ values. Using equidistant
$\lambda$ values, the estimates obtained from the cubic spline and trapezoidal
quadrature are both heavily biased. As the number of $\lambda$ values
increases, the accuracy and precision improve. Using eleven equidistant
$\lambda$ values, for example, the Lagrange and Newton polynomials give the
most accurate estimates of $\Delta F$ but with large uncertainty compared to
that of the cubic spline and trapezoidal quadrature. The estimates obtained
from cubic spline are slightly biased but with a low uncertainty. Using
non-equidistant $\lambda$ values, the Lagrange and Newton polynomials are able
to most accurately estimate $\Delta F$. These results suggest that using
non-equidistant $\lambda$ values is preferred to equidistant $\lambda$ values.
Quadrature results, however, do not improve with the use of non-equidistant
$\lambda$ values. We also note that, unlike quadrature, the steep free energy
curvature for system two does not appear to significantly decrease the accuracy
of $\Delta F$ estimates obtained from the Lagrange and Newton polynomials.
These results are consistent with our previous regression study \cite{shyu09}.

\begin{table}[tbp]
\centering \footnotesize
\begin{tabular}{r r r r r}
\multicolumn{5}{c}{ } \\
\multicolumn{5}{l}{System One \cite{shyu09}} \\
\hline
 & \multicolumn{1}{c}{Six Equid.} & \multicolumn{1}{c}{Six Non-Equid.} &
 \multicolumn{1}{c}{Eleven Equid.} & \multicolumn{1}{c}{Eleven Non-Equid.} \\
\hline
Trapezoidal Rule & 1.2496 (0.0210) & 1.1390 (0.0212) & 0.3270 (0.0152) &
0.3073 (0.0150) \\
Cubic Spline & 0.4686 (0.0198) & -0.1138 (0.0218) & 0.0752 (0.0152) &
-0.0047 (0.0152) \\
Lagrange/Newton & 0.1606 (0.0200) & -0.0208 (0.0212) & 0.0034 (0.0377) &
0.0006 (0.0151) \\
\hline
\multicolumn{5}{c}{} \\
\multicolumn{5}{l}{System Two \cite{shyu09}}\\
\hline
 & \multicolumn{1}{c}{Six Equid.} & \multicolumn{1}{c}{Six Non-Equid.} &
 \multicolumn{1}{c}{Eleven Equid.} & \multicolumn{1}{c}{Eleven Non-Equid.} \\
\hline
Trapezoidal Rule & -1.8944 (0.0232) & -1.4944 (0.0231) & -0.5077 (0.0161) &
-0.4094 (0.0154) \\
Cubic Spline & -0.8228 (0.0212) & 0.2035 (0.0234) & -0.1424 (0.0159) &
0.0093 (0.0156) \\
Lagrange/Newton & -0.3495 (0.0210) & 0.0467 (0.0227) & -0.0053 (0.0335) &
-0.0006 (0.0154) \\
\hline
\end{tabular}
\caption[Averages and standard deviations of the biases ($\Delta F
_{\textrm{exact}} - \Delta F _{\textrm{estimate}}$) for system one and two
using six and eleven equidistant and non-equidistant $\lambda$ values]{Averages
and standard deviations of the biases ($\Delta F _{\textrm{exact}} - \Delta F
_{\textrm{estimate}}$) for system one and two using six and eleven equidistant
and non-equidistant $\lambda$ values. The exact solution for system one is
$\Delta F ^{\textrm{one}} _{\textrm{exact}} = - \frac{1}{2} \ln \frac{1}{4}$
and for system two $\Delta F ^{\textrm{two}} _{\textrm{exact}} = - \frac{1}{2}
\ln 5$ as previously reported by Shyu and Ytreberg \cite{shyu09}. Standard
deviations of the biases are shown in parentheses.} \label{tab:analytical}
\end{table}

Overall, the Lagrange and Newton polynomials provide the most accurate
estimates, but often with larger uncertainty than that of spline or quadrature.
Results from additional cursory simulations suggest that the variation of the
$\Delta F$ estimates are predominately caused by the oscillations from the use
of high degree polynomials (data not shown). Use of non-equidistant $\lambda$
values significantly improves the numerical stability and increases the
accuracy and precision of $\Delta F$ estimates when using polynomial
interpolation. The use of non-equidistant $\lambda$ values, however, provides
only slight improvement of the estimates obtained from the trapezoidal
quadrature compared to that of equidistant. The results also show that the
$\Delta F$ estimates obtained from spline interpolation are more accurate than
that of trapezoidal rule. The estimates from quadrature are all heavily biased.
The estimates from the cubic spline have similar uncertainty but are not as
accurate as that of Lagrange and Newton. Thus, we conclude that use of both
polynomial and spline interpolation techniques with non-equidistant $\lambda$
values provide accurate estimates of $\Delta F$.

%
\subsection{BUQ Solvation}
\begin{figure}[tbp]
\centering \footnotesize
\begin{tabular}{c c}
\includegraphics[width=3.0in]{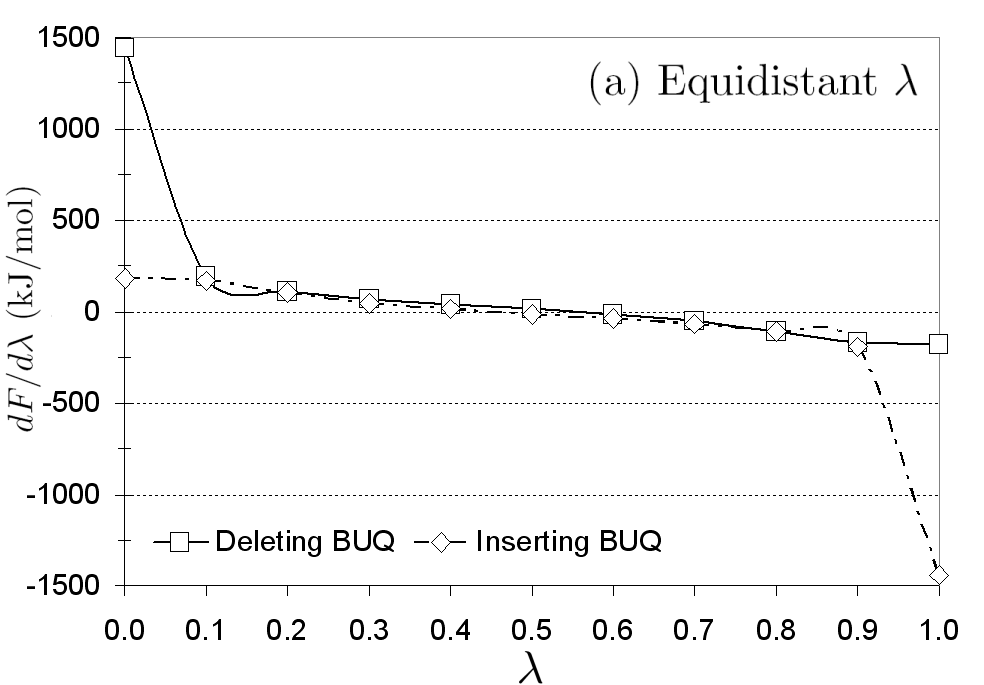} &
\includegraphics[width=3.0in]{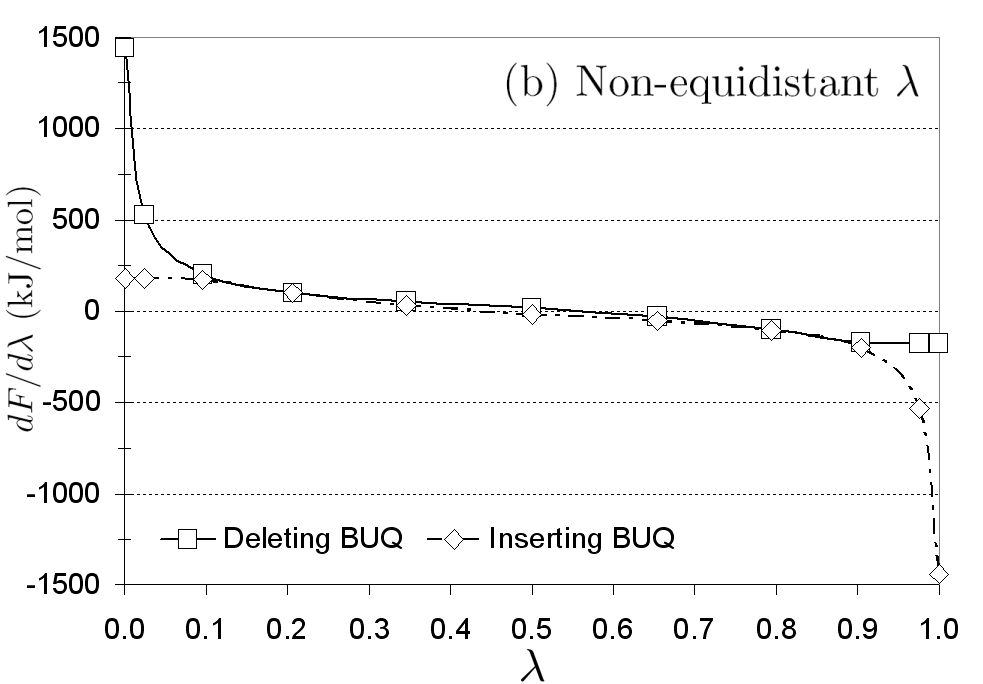} \\
\end{tabular}
\caption[Solvation free energy curves for BUQ with eleven equidistant (a) and
non-equidistant (b) $\lambda$ values]{Solvation free energy curves for BUQ with
eleven equidistant (a) and non-equidistant (b) $\lambda$ values. Both insertion
(solid curve) and deletion (dotted curve) were computed. The fitting curves
shown were constructed using Lagrange interpolating polynomials. Note that the
endpoint singularity problem is present when the growing and vanishing atoms
are too close to the other atoms ($\lambda = 0.0$ or 1.0). The use of
non-equidistant $\lambda$ values allows a smoother polynomial fit.}
\label{fig:buq_all}
\end{figure}

As detailed above, we first performed a two-stage simulation for Lennard-Jones
and partial charges in order to obtain the reference solvation free energies
$\Delta F ^{\textrm{ins}} _{\textrm{ref}}$ and $\Delta F ^{\textrm{del}}
_{\textrm{ref}}$. We then determined the free energy estimate with no soft-core
scaling utilizing a single stage, i.e., Lennard-Jones and partial charges were
simultaneously changed. Figure \ref{fig:buq_all} shows the free energy curves
from the simulations using eleven equidistant (b) and non-equidistant $\lambda$
values (a). The curves are all well-behaved except when $\lambda$ is close to
either 0.0 or 1.0. As expected, the figures show clear evidence of the endpoint
singularity when the vanishing or growing atoms become close to the other
atoms. The statistical uncertainty progressively grows larger as $\lambda$
approaches 0.0 or 1.0. The rapid changes on the free energy curves at the
endpoint introduce significant errors into the numerical integration using
trapezoid quadrature. It is worth noting that the use of non-equidistant
$\lambda$ significantly alleviates the endpoint singularity problem, even for
quadrature, because more data points are used at each end.

\begin{table}[tbp]
\centering \footnotesize
\begin{tabular}{r r r r r}
\multicolumn{5}{c}{ } \\
\hline
 & Del. Equid. & Del. Non-Equid. & Ins. Equid. & Ins. Non-Equid. \\
\hline
Trapezoidal Rule & -40.2 & -9.5 & 39.6 & 9.5 \\
Cubic Spline & -26.8 & 1.0 & 27.0 & -1.1 \\
Lagrange/Newton & -14.0 & -0.4 & 14.5 & 0.4 \\
Regression \cite{shyu09} & -14.3 & -0.4 & 14.5 & 0.3 \\
\hline
\end{tabular}
\caption[Free energy biases ($\Delta F _{\textrm{ref}} - \Delta F
_{\textrm{estimate}}$) using eleven equidistant and non-equidistant $\lambda$
values]{Free energy biases ($\Delta F _{\textrm{ref}} - \Delta F
_{\textrm{estimate}}$) using eleven equidistant and non-equidistant $\lambda$
values. The reference free energy estimates ($\Delta F ^{\textrm{del}}
_{\textrm{ref}}$ and $\Delta F ^{\textrm{ins}} _{\textrm{ref}}$) were obtained
from a two-stage simulation using soft-core scaling and are utilized as
reference values to evaluate the polynomial techniques. The free energy for BUQ
deletion is $\Delta F ^{\textrm{del}} _{\textrm{ref}} = 31.6$ kJ/mol and
insertion $\Delta F ^{\textrm{ins}} _{\textrm{ref}} = -31.5$ kJ/mol.}
\label{tab:summary}
\end{table}

Table \ref{tab:summary} summarizes the $\Delta F$ estimates obtained using both
interpolation and regression techniques with equidistant (Table
\ref{tab:summary}A) and non-equidistant $\lambda$ values (Table
\ref{tab:summary}B). The reference free energies $\Delta F ^{\textrm{ins}}
_{\textrm{ref}}$ and $\Delta F ^{\textrm{ins}} _{\textrm{ref}}$ were used to
estimate the accuracy and precision of these polynomial fitting techniques. For
equidistant $\lambda$ values, while none of the integration techniques give an
accurate estimate of $\Delta F$, Lagrange and Newton provide the best results.
The use of non-equidistant $\lambda$ values, significantly improves the
accuracy of all $\Delta F$ estimates. The improvement is particularly evident
with the estimates obtained using Lagrange and Newton polynomials. Cubic spline
interpolation delivers slightly biased estimates but is still considerably more
accurate than that of trapezoidal quadrature.

For BUQ solvation, our study has shown that using the polynomial fitting
techniques for interpolation and extrapolation and use of non-equidistant
$\lambda$ values, $\Delta F$ can be estimated within the accuracy of $\pm 0.4$
kJ/mol \emph{without using soft-core scaling and performing separate
simulations for Lennard-Jones or partial charges}. Steinbrecher et al.
\cite{steinbrecher07} showed that the soft-core potentials are most promising
for alchemical TI simulations and deliver $\Delta F$ estimates with accuracies
of 0.1 kcal/mol. However, our simulations suggest that similar accuracy can be
achieved using non-equidistant $\lambda$ values and polynomial fitting
techniques without soft-core scaling.

%
\section{Conclusion}
We have implemented polynomial interpolation and regression techniques to
estimate a small molecule solvation free energy using thermodynamic integration
(TI) simulation data. Our study utilized two test systems with analytical
$\Delta F$ values to objectively quantify the accuracy and precision of the
interpolation techniques. We also used both interpolation and regression to
estimate the solvation free energy for BUQ. These polynomial approaches rely on
constructing globally optimal polynomials that best fit the TI data, which are
then used to estimate $\Delta F$. Additional algebraic calculations are done to
permit analytical integration eliminating potential errors due to numerical
evaluation on the integral.

Studies have shown that the use of soft-core scaling improves the accuracy of
$\Delta F$ estimates within $\pm 0.1$ kJ/mol for alchemical TI simulations
\cite{steinbrecher07}. Our simulations, however, suggest that similar accuracy
can also be achieved using non-equidistant $\lambda$ and polynomial
interpolation or regression techniques without soft-core scaling.

We have shown that polynomial and spline interpolation techniques provide more
accurate and precise $\Delta F$ estimates than trapezoidal quadrature for the
systems studied here. However, we caution that polynomial interpolation
techniques possess some inherent weakness. The most significant is that use of
more than eleven $\lambda$ produces interpolation polynomials that are
numerically unstable. As a consequence, our approach is limited to use of
eleven or less equidistant or non-equidistant $\lambda$ values or one may
judicially select subsets of the $\lambda$ data that best represent the
underlying functional form of the free energy curve.

Our results also illustrate the importance of selection of $\lambda$ values if
one wishes to use polynomial interpolation techniques to improve the accuracy
of $\Delta F$ estimates. For the systems studied here, use of non-equidistant
$\lambda$ always produced more accurate $\Delta F$ estimates than use of
equidistant $\lambda$. Using twelve or less $\lambda$ values the Lagrange and
Newton polynomials always produced the highest accuracy. The $\Delta F$
estimates via cubic spline were more accurate than quadrature but less accurate
than Lagrange and Newton. We note that for data containing more twelve
$\lambda$ values, cubic spline is expected to produce more accurate results
than Lagrange and Newton because of superior numerical stability.

To better facilitate the use of polynomial fitting techniques, we propose
several guidelines for estimating $\Delta F$. Researchers who already have very
well converged TI data with eleven or less $\lambda$ values are encouraged to
utilize the Lagrange or Newton polynomial interpolation approach. For
researchers who already have data from equilibrium simulations with twelve or
more $\lambda$ values, the spline interpolation technique should be used
instead. Researchers that have not yet generated data are strongly encouraged
to use non-equidistant $\lambda$ values and polynomial fitting techniques.
Finally, if the TI data is noisy or not very well converged then regression
\cite{shyu09} should be used instead of interpolation. To allow researchers to
immediately utilize these methods, free software and documentation is provided
via \url{http://www.phys.uidaho.edu/ytreberg/software}.

%
\section*{Acknowledgements}
This research was supported by Idaho NSF-EPSCoR, Bionanoscience at University
of Idaho (BANTech), and the Initiative for Bioinformatics and Evolutionary
Studies (IBEST) at University of Idaho.

%
\singlespace
\bibliographystyle{plain}

\end{document}